# A Ni-Fe Layered Double Hydroxide-Carbon Nanotube Complex for Water Oxidation

Ming Gong[1†], Yanguang Li[1†], Hailiang Wang[1], Yongye Liang[1], Justin Zachary Wu[1], Jigang Zhou[2], Jian Wang[2], Tom Regier[2], Fei Wei[3] and Hongjie Dai[1]*

[1]Department of Chemistry, Stanford University, Stanford, CA 94305, USA

[2]Canadian Light Source Inc., Saskatoon, SK S7N 0X4, Canada.

[3]Department of Chemical Engineering, Tsinghua University, Beijing, 100084, China

[†]These authors contribute equally

*Correspondence to: hdai@stanford.edu

# Abstract

**Highly active, durable and cost-effective electrocatalysts for water oxidation to evolve oxygen gas hold a key to a range of renewable energy solutions including water splitting and rechargeable metal-air batteries. Here, we report the synthesis of ultrathin nickel iron layered double hydroxide (NiFe-LDH) nanoplates on mildly oxidized multi-walled carbon nanotubes. Incorporation of Fe into the nickel hydroxide induced the formation of NiFe layered double hydroxide. The crystalline NiFe-LDH phase in nanoplate form is found to be highly active for oxygen evolution reaction (OER) in alkaline solutions. For NiFe-LDH grown on a network of carbon nanotubes, the resulting NiFe-LDH/carbon nanotube complex exhibit higher electro-catalytic activity and stability for oxygen evolution than commercial precious metal Ir catalysts.**

## Introduction

Oxygen electrochemistry has been intensely researched in the pursuit of sustainable and efficient energy conversion and storage solutions. Oxygen evolution reaction (OER) is the process of generating molecular oxygen through electrochemical oxidation of water, and holds a key to a number of important energy conversion and storage processes such as water splitting and rechargeable metal-air batteries[1-7]. OER proceeds through multistep proton-coupled electron transfer, and is kinetically sluggish[8,9]. An effective electrocatalyst is needed in order to expedite the reaction, reduce the overpotential and thus enhance the energy conversion efficiency. Currently, the most active OER catalysts are $IrO_2$ or $RuO_2$ in acidic or alkaline solutions[4,10], but these catalysts suffer from the scarcity and high cost of precious metals. Extensive efforts have been taken to develop highly active, durable and low cost alternatives such as first-row transition metal oxides[11-15] and perovskites[16,17]. However, most non-precious metal catalysts developed still underperform the Ir benchmark.

Here, we report a NiFe layered double hydroxide-carbon nanotube complex (NiFe-LDH/CNT) with higher OER catalytic activity and stability than commercial Ir based catalysts. The key aspects of this catalyst are the formation of ultra-thin nanoplates of a highly OER active NiFe layered double hydroxide (LDH) structure (the same structure as alpha-phase Ni hydroxide (α-phase $Ni(OH)_2$) and association of the nanoplates with CNTs that can form interconnected electrically conducting networks. Even though electrodeposited NiFe mixed oxide/hydroxide have been made for OER previously[18-21], this was the first time that crystalline NiFe-layered double hydroxide was synthesized chemically to obtain highly active electrocatalysts for the oxygen evolution reaction in alkaline media. The turn-over frequency of the NiFe-LDH/CNT catalyst exceeded any previous Ni-Fe compounds[18,22-25] and was comparable to the most active perovskite based catalyst[17].

**Results**

A three-step process was developed to synthesize the NiFe-LDH/CNT complex. Nickel acetate and iron nitrate (with a molar ratio of Ni/Fe = 5) were hydrolyzed and selectively coated onto mildly oxidized multi-walled carbon nanotubes (by a modified Hummers' method with 1/3 amount of $KMnO_4$, see Supporting Information for details) in a N,N-dimethylformamide (DMF) solution at 85 °C. The intermediate product was re-dispersed in a $DMF/H_2O$ mixed solvent and then subjected to a solvothermal treatment at 120 °C for 12 h, followed by a second solvothermal step at 160 °C for 2 h. The solvothermal treatment led to the crystallization of nanoplates and partial reduction of the oxidized CNTs.

The size, morphology and structure of the resulting material were characterized by scanning electron microscopy (SEM) and transmission electron microscopy (TEM). SEM (Figure 1a) showed that ultra-thin plates (reflected by poor image contrast) were grown over a CNT network. TEM (Figure 1b) showed that the nanoplate size was typically around 50 nm and nearly transparent to electron beams due to the ultra thin nature. We also observed 3-5 nm nanoparticles with higher contrast along the edges of nanoplates and the outer-walls of CNTs. High resolution TEM (HRTEM) (Figure 1c) of these nanoparticles revealed a different set of lattice fringes corresponding to iron oxide ($FeO_x$). Spatially resolved energy dispersive spectroscopy (EDS) analysis (spatial resolution of ~2 nm) also suggested small amount of $FeO_x$ nanoparticles distributed over the nanoplate/CNT complex (Figure S1).

X-ray diffraction (XRD) pattern (Figure 1d) of the final product was consistent with the α-phase $Ni(OH)_2$ (the same as LDH) with a greater interlayer distance compared to β-phase

Ni(OH)$_2$. From the width of (003) and (006) diffraction peaks, the thickness of nanoplates was estimated to be ~5 nm, confirmed by atomic force microscopy (AFM) topological height analysis (Figure S2). Interestingly, pure β-phase Ni(OH)$_2$ nanoplates on CNT networks were synthesized by the same method in the absence of any iron precursors. The drastically different phases suggested that incorporation of a Fe precursor to Ni induced the formation of NiFe hydroxide in LDH structure. It was known that Fe$^{3+}$ could replace Ni$^{2+}$ in the Ni(OH)$_2$ lattice, forming a stable LDH structure[26-30]. The excessive cationic charge due to Fe$^{3+}$ was balanced by anion intercalation between the hydroxide layers[28]. X-ray photoelectron spectroscopy (XPS, Figure S3) corroborated the existence of both Fe and Ni in the hybrid material. The Fe species was found to be mostly in the +3 oxidation state from high resolution Fe 2p spectrum (Figure S3d). X-ray absorption near edge structure (XANES) measurements confirmed the Ni$^{+2}$ oxidation state and Fe$^{+3}$ oxidation state (Figure S4a and Figure S4b). Both Fe and Ni signals were detected on nanoplates in areas free of decorating FeO$_x$ nanoparticles by spatially resolved EDS spectroscopy (Figure S1). These results suggested the synthesis of ultra-thin NiFe-LDH nanoplates decorated with FeO$_x$ nanoparticles grown over a network of gently oxidized multi-walled carbon nanotubes (Figure 1).

We investigated the electrocatalytic OER activity of NiFe-LDH/CNT in alkaline solutions (0.1 M KOH and 1 M KOH) in a standard three electrode system (see details in Supporting Information). The material was uniformly casted to a glassy carbon electrode (loading ~0.2 mg/cm$^2$) for recording iR-corrected OER polarization curves at a slow scan rate of 5 mV/s to minimize capacitive current (Figure S5). During the measurements, the working electrode was continuously rotating at 1600 rpm to remove the generated oxygen bubbles. For comparison, a commercial Ir/C catalyst (20 wt% Ir on Vulcan carbon black from Premetek Co. with the same

~0.2 mg/cm$^2$ loading) was measured side by side. In 0.1 M KOH, the anodic current recorded with the NiFe-LDH/CNT catalyst showed a sharp onset of OER current at ~1.50 V versus the reversible hydrogen electrode (RHE) (Figure 2a). The Ir/C catalyst afforded similar onset potential, but its OER current density fell below our NiFe-LDH/CNT hybrid at ~1.52 V. In 1 M KOH, the OER onset potential of the NiFe-LDH/CNT reduced considerably to ~1.45 V versus RHE (Figure 2a). We also loaded and tested the catalysts on carbon fiber paper (loading density ~ 0.25 mg/cm$^2$, effective area 1 cm$^2$) and observed similar activity trend and pH dependence (Figure 2b) as on rotating disk electrode. The peak around 1.43 V of NiFe-LDH/CNT in 1 M KOH is assigned to Ni(II)/Ni(III or IV) redox process[31]. The NiFe-LDH/CNT electrocatalyst was among the most active non-precious metal electrocatalysts[11-15,24,32] (Table S1).

We calculated a high turnover frequency (TOF) of 0.56 s$^{-1}$ associated with NiFe-LDH/CNT at an overpotential of 300 mV in 1 M KOH assuming all the metal sites were involved in the electrochemical reaction (see Supporting Information). This value represented the lower limit since some of these metal sites in the nanoplates were electrochemically non-accessible. Even in this case, the TOF of our NiFe-LDH/CNT catalyst was still about three times higher than previous mixed nickel and iron oxide electrocatalysts (the previous highest TOF was 0.21 s$^{-1}$ at similar experimental conditions)[18,22-25]. Our result here also compared favorably to the TOF of a high performance perovskite material (~0.6 s$^{-1}$ calculated based on only surface active sites)[17].

Besides high OER activity, the NiFe-LDH/CNT catalyst exhibited good durability in alkaline solutions (Figure 2c and 2d). When biased galvanostatically at 5 mA/cm$^2$ on carbon fiber paper electrodes, the operating potential of the catalyst stayed nearly constant at ~1.52 V (corresponding to an overpotential of 0.29 V) in 0.1 M KOH, whereas the Ir/C catalyst showed an increase in overpotential by ~ 20 mV (Figure 2d). In 1 M KOH (Figure 2d), the working potential of NiFe-LDH/CNT was lowered to ~1.48 V to deliver a 5 mA/cm$^2$ current density and the catalyst was also more stable than the commercial Ir catalyst. Consistent OER durability data were also recorded on rotating disk electrodes (Figure 2c).

We fitted the polarization curves obtained with the NiFe-LDH/CNT hybrid on carbon fiber paper electrodes at various pHs to the Tafel equation ($\eta = b\log(j/j_0)$, where $\eta$ is the overpotential, $b$ is the Tafel slope, $j$ is the current density and $j_0$ is the exchange current density)[33]. The NiFe-LDH/CNT catalyst exhibited a Tafel slope of $b \sim 35$ mV/decade in 0.1 M KOH and $b \sim 31$ mV/decade in 1 M KOH (Figure 3a). This value was smaller than that of Ir/C reference (~40 mV/decade)[33]. Consistently, the NiFe-LDH/CNT catalyst exhibited higher activity and stability than Ir/C at various current densities of 10 mA/cm$^2$, 20 mA/cm$^2$ (~50 mV and ~60 mV difference in overpotential after 1 hour operation, Figure 3b). Similar trend of OER activity and stability were observed under maximal loading of the hybrid catalysts on CFP (Figure S6).

To confirm oxygen evolution, we carried out high-current and long time electrolysis by passing through more charges than would be needed for completely oxidizing carbon in the NiFe-LDH/CNT catalysts (Figure S7). High OER activities remained and CNT structures were observed in the catalyst after the long OER operation when examined by SEM and TEM (Figure S7c and Figure S7d), suggesting that the CNTs in the catalyst complex were not removed by oxidative etching under the OER conditions used. Note that pure CNTs without NiFe-LDH showed negligible current in the < 1.5 V voltage range (Figure S8 green curve) over which OER was highly active for the NiFe-LDH/CNT catalyst (Figure S8 black curve). Clearly, the high anodic current density observed with the NiFe-LDH/CNT catalyst was dominant by OER catalyzed by the hybrid material rather than oxidative corrosion of CNTs. We analyzed the gaseous products from the OER of NiFe-LDH/CNT catalysts on non-carbon based electrode using gas chromatography, and found that oxygen was the only product, with a Faradaic efficiency similar to that of an $IrO_2$ benchmark [$IrO_2$ nanoparticle (from Premetek Co.)] (Figure S9).

## Discussion

Our electrochemical data suggested that the NiFe-LDH/CNT complex was a novel electrocatalyst material with high OER activity and stability in basic solutions. The high electrocatalytic performance was mainly attributed to the NiFe-LDH phase. Strong association of the LDH with CNTs further facilitated charge transport and improved the catalyst (Figure 4b). In a series of control experiments, we observed that the NiFe LDH material alone without CNTs showed high activity for OER catalysis and was superior to amorphous NiFe oxide formed by electrodeposition[18] on glass carbon electrodes (Figure 4a). The NiFe-LDH phase was also more active for OER than a mixture of $\beta$-$Ni(OH)_2$ and $FeO_x$ nanoparticles, with and without association with CNTs (Figure 4b). When loaded into Ni foams to facilitate better contact and charge transport by the Ni foam substrate, the NiFe-LDH nanoplates alone without any carbon additive exhibited high activity and stability over several days of OER operation at 20 mA/cm$^2$ in 1 M KOH (Figure S10). We also performed XRD analysis of the NiFe-LDH/CNT catalysts after OER operation for 3 hours. The XRD pattern was consistent with a crystalline NiFe layered double hydroxide phase (Figure S7e), suggesting no change in the LDH phase through OER catalysis. Thus, we chemically synthesized a crystalline NiFe-LDH phase as a highly OER active material in basic solutions.

Interaction between NiFe-LDH and CNTs afforded by direct nucleation and growth of LDH nanoplates on the functional groups on CNTs contributed to the optimal OER activity of the NiFe-LDH/CNT complexes. XANES measurements were employed to reveal the interactions (Figure 4c). At carbon K edge absorption, the NiFe-LDH/CNT showed a drastic increase of carbonyl π* peak intensity at around ~288.5 eV[34] compared to the CNT control without any NiFe-LDH plates. This was attributed to the formation of M-O-C (M=Ni, Fe) bonding via the carboxyl group, leading to large perturbations to the carbon atoms in the carbonyl groups. The drastic changes in the X-ray spectroscopy revealed strong interaction effects of NiFe-LDH nanoplates and CNTs, which facilitated charge transport and favored high OER activity and stability. Accordingly, NiFe-LDH/CNT hybrid material exhibited higher OER activity than NiFe-LDH nanoplate alone, NiFe-LDH mixed with carbon black and NiFe-LDH mixed with CNT (Figure 4b and Figure S11).

In conclusion, we devised a strategy for the nucleation and growth of NiFe-LDH nanoplates on mildly oxidized CNTs. We uncovered the high intrinsic OER electrocatalytic activity of the crystalline NiFe-LDH phase and observed the underlying CNT network enhancing electron transport and facilitating high OER activity of the NiFe-LDH nanoplate/CNT complex. This led to an electrocatalyst that outperformed Ir in both activity and stability in basic solutions, opening a new venue to advanced, low cost oxygen evolution electrocatalysts for energy applications.

## Author Contributions

M. G., Y. Li and H.D. conceived the project and designed the experiments. M. G., Y. Li, H.W., Y. Liang and J. W. performed material synthesis, structural characterization and electrochemical measurements. J.Z., J.W. and T.R. performed the XANES measurement and analysis. F.W. synthesized carbon nanotubes. M. G., Y. Li, H. W., Y. Liang, J. W., J. Z., J. W., T. R. and H.D. analysed the data. M. G., Y. Li and H.D. co-wrote the paper. All authors discussed the results and commented on the manuscript.

## Additional information

The authors declare no competing financial interests.

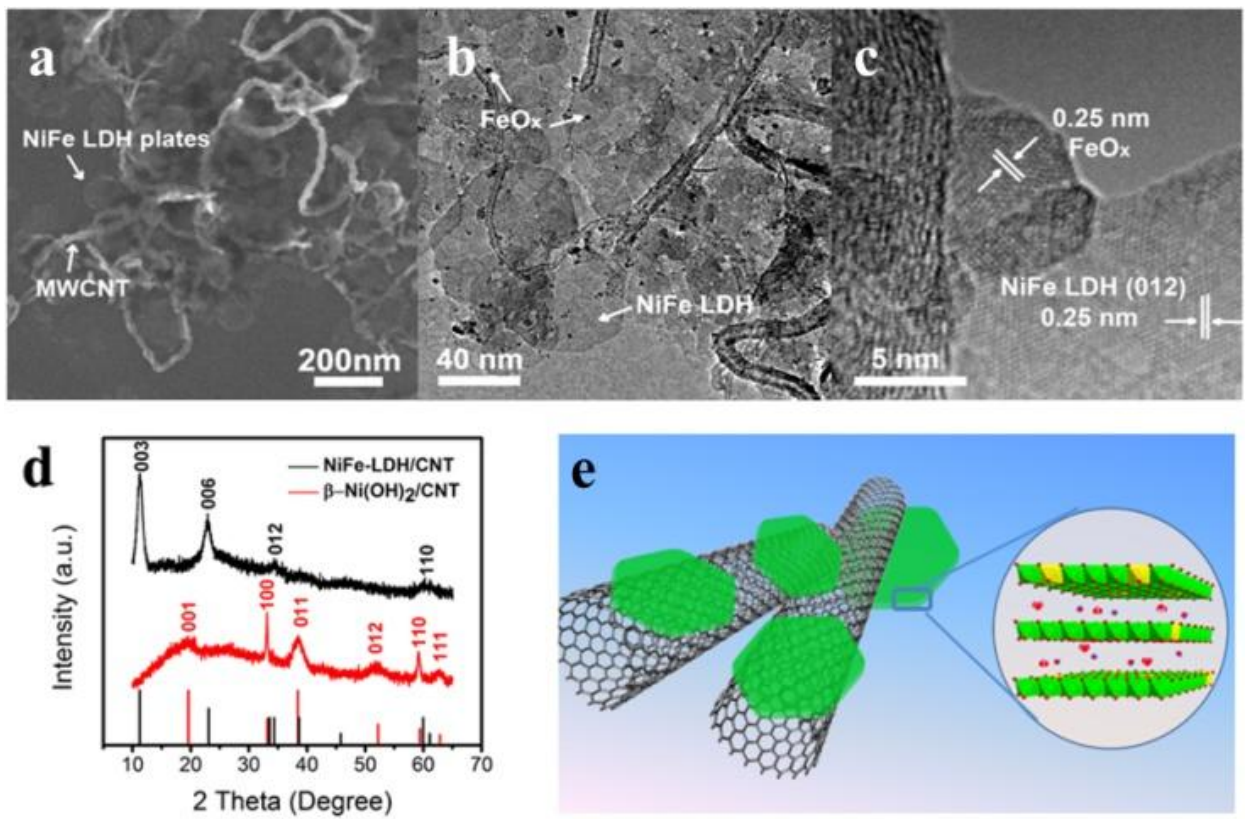

**Figure 1.** Ultrathin NiFe layered double hydroxide nanoplates grown on carbon nanotubes. (a) SEM image of NiFe-LDH nanoplates grown over a network of mildly oxidized MWCNTs. (b-c) TEM images of the NiFe-LDH/CNT hybrid. Arrows point to individual NiFe-LDH plates and smaller iron oxide particles. (d) XRD spectra of NiFe-LDH/CNT (black) and a control β-Ni(OH)$_2$/CNT sample (red, synthesized without the iron precursor). The lines correspond to standard XRD patterns of α-Ni(OH)$_2$ (black, JCPDS card No. 38-0715) and β-Ni(OH)$_2$ (red, JCPDS card No. 14-0117) (e) Schematic showing the hybrid architecture and layered double hydroxide crystal structure.

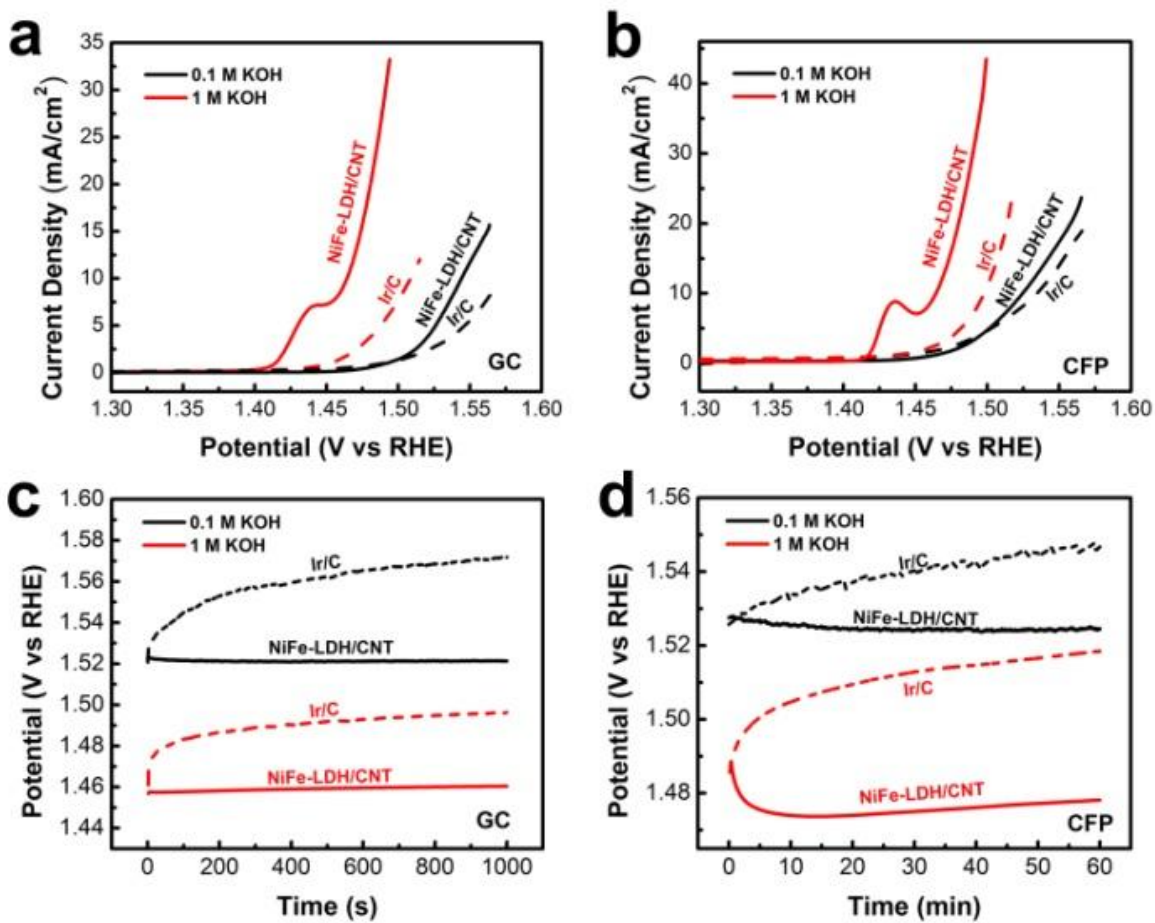

**Figure 2.** Electrochemical performance of NiFe-LDH/CNT hybrid OER catalyst. (a) iR-corrected polarization curves of NiFe-LDH/CNT hybrid and Ir/C catalysts on glassy carbon (GC) electrode in 0.1 M KOH and 1 M KOH respectively, measured with a catalyst loading of 0.2 mg/cm$^2$ for both NiFe-LDH/CNT and Ir/C at a continuous electrode rotating speed of 1600 rpm. (b) iR-corrected polarization curves of NiFe-LDH/CNT hybrid and Ir/C catalysts on carbon fiber paper (CFP), measured with a catalyst loading of 0.25 mg/cm$^2$. See Supporting Information and Figure S5 for details of iR correction. (c) Chronopotentiometry curves of NiFe-LDH/CNT hybrid and Ir/C catalyst on GC at a constant current density of 2.5 mA/cm$^2$. (d) Chronopotentiometry curves of NiFe-LDH/CNT hybrid and Ir/C catalyst on CFP at a constant current density of 5 mA/cm$^2$.

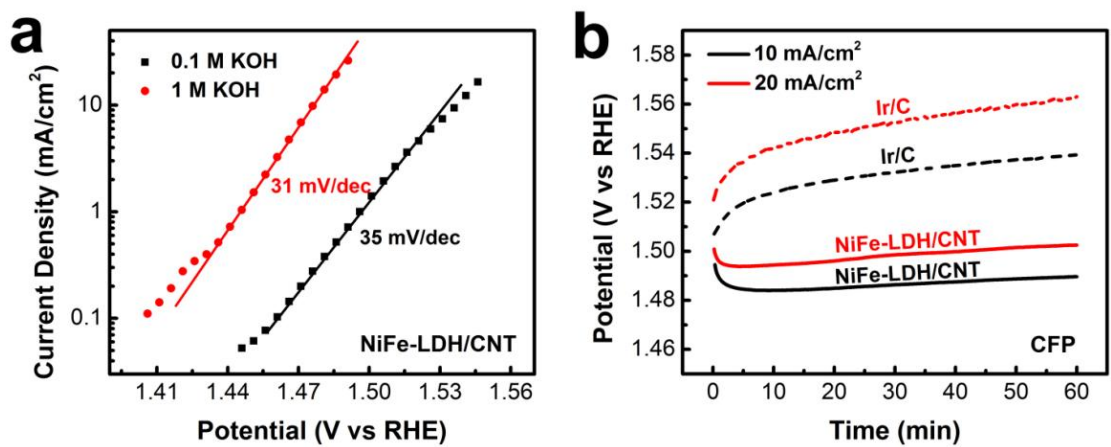

**Figure 3**. (a) Tafel plots of NiFe-LDH/CNT catalyst loaded on CFP (at a loading of 0.25 mg/cm$^2$) recorded at 0.1 mV/s in 0.1 M KOH and 1 M KOH. (b) Chronopotentiometry curves of NiFe-LDH/CNT hybrid and Ir/C catalyst on CFP (at a loading of 0.25 mg/cm$^2$) under higher current densities of 10 mA/cm$^2$ and 20 mA/cm$^2$

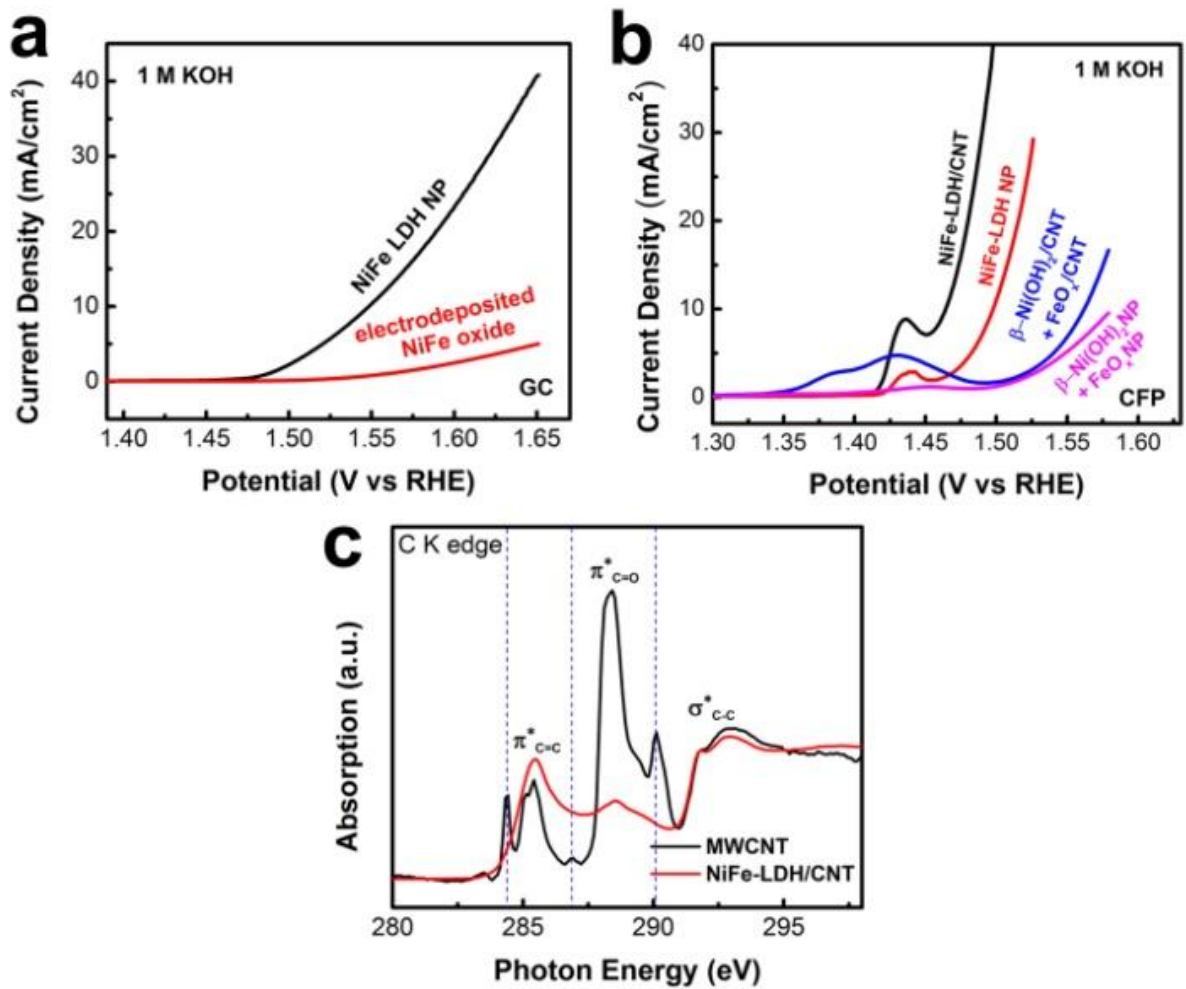

**Figure 4.** (a) Polarization curves of free NiFe LDH nanoplates (loading of 0.2 mg/cm$^2$) without any CNTs and electrodeposited NiFe oxide [cathodically electrodeposited under the current density of 1 mA/cm$^2$ for 600 second 0.09 M Ni(NO$_3$)$_2$ and 0.01 M Fe(NO$_3$)$_3$ solution (pH = 2) (loading of 2.4 mg/cm$^2$)[16]] in 1 M KOH on glassy carbon electrodes. (b) Polarization curves of NiFe-LDH/CNT hybrid, NiFe-LDH plate alone, a physical mixture of β-Ni(OH)$_2$/CNT and FeO$_x$/CNT and a physical mixture of β-Ni(OH)$_2$ and FeO$_x$ nanoparticles loaded on CFP (with a loading of 0.25 mg/cm$^2$) in 1 M KOH. (c) C K-edge XANES spectra of NiFe-LDH/CNT (black) and pure multi-walled carbon nanotube (MWCNT, red) without coupling to LDH. The blue dashed lines mark defects at 284.4 eV, σ*$_{C-H}$ at 286.9 eV and π*$_{O-C(O)-O}$ (carbonate) at 290.1 eV.

Supporting Information for

# An Advanced Ni-Fe Layered Double Hydroxide Electrocatalyst for Water Oxidation

Ming Gong[†], Yanguang Li[†], Hailiang Wang, Yongye Liang, Justin Z. Wu, Jigang Zhou, Jian Wang, Tom Regier, Fei Wei and Hongjie Dai*

## Experiment Details

Oxidization of multi-walled carbon nanotubes (MWCNTs): MWCNTs were oxidized by a modified Hummers method[35, 36]. 1g of MWCNTs (FloTube 9000 from CNano Technology Ltd.) were first purified by calcination at 500 °C for 1 h and washed with 70 ml of diluted hydrochloric acid (10 wt%) to remove metal residues. The products were then filtered, washed and lyophilized. After that, 23 ml of concentrated sulfuric acid was added to the purified MWCNTs in a 250 ml round-bottom flask, and the mixture was stirred at room temperature overnight. Next, the solution was transferred to an oil bath. After its temperature was raised to 40 °C, 350 mg of $NaNO_3$ was added, followed by the slow addition of 1 g of $KMnO_4$ while keeping the reaction temperature below 45°C. The solution was kept stirring at 40°C for 30 min. And then, 3 ml of water was added to the flask, followed by another 3 ml after 5 minutes. After another 5 minutes, 40 ml of water was added. 15 minutes later, the flask was removed from the oil bath and 140 ml of water and 10 ml of 30% $H_2O_2$ were added to end the reaction. Oxidized MWCNTs were collected, repetitively washed with 5% HCl solution and then water, and finally lyophilized.

Synthesis of NiFe hydroxide carbon nanotube hybrid (NiFe-LDH/CNT): In a typical synthesis, ~2 mg of oxidized MWCNTs were dispersed in 4 ml of anhydrous N,N-dimethylformamide (DMF) assisted by sonication for 10 min. After that, 400 μl of 0.2 M nickel acetate ($Ni(OAc)_2$) and 80 μl of 0.2 M ferrous nitrate ($Fe(NO_3)_3$) aqueous solution were added. The solution was vigorously stirred at 85 °C for 4 hours. Then, the product was collected and dispersed in a mixture of ~4ml DMF and ~8 ml of water, and transferred to a 40 ml Teflon lined stainless steel autoclave for solvothermal reaction at 120 °C for 12 hours, followed by another solvothermal treatment at 160 °C for 2 hours. The final product was collected by centrifuge, repetitively washed with water and lyophilized.

Synthesis of control groups of NiFe-LDH nanoplate, Nickel hydroxide (Ni(OH)$_2$) nanoplate, Iron oxide (FeO$_x$) nanoparticle, β-Nickel hydroxide nanoplate carbon nanotube hybrid (β-Ni(OH)$_2$/CNT), Iron oxide nanoparticle carbon nanotube hybrid (FeO$_x$/CNT): All controls except hydrothermally treated NiFe-LDH/CNT were synthesized by similar procedures of NiFe-LDH/CNT synthesis with different precursors. NiFe-LDH nanoplate was synthesized with 400 μl of 0.2 M Ni(OAc)$_2$ and 80 μl of 0.2 M (Fe(NO$_3$)$_3$) aqueous solution (without oxidized MWCNTs precursor). Ni(OH)$_2$ nanoplate was synthesized with 400 μl of 0.2 M Ni(OAc)$_2$ aqueous solution (without oxidized MWCNTs and Fe(NO$_3$)$_3$ precursors). FeO$_x$ nanoparticle was synthesized with 400 μl of 0.2 M Fe(NO$_3$)$_3$ aqueous solution (without oxidized MWCNTs and Ni(OAc)$_2$ precursors). β-Ni(OH)$_2$/CNT was synthesized with ~2 mg of oxidized MWCNTs and 400 μl of 0.2 M Ni(OAc)$_2$ aqueous solution (without Fe(NO$_3$)$_3$ precursors). FeO$_x$/CNT was synthesized with ~2 mg of oxidized MWCNTs and 400 μl of 0.2 M Fe(NO$_3$)$_3$ aqueous solution (without Ni(OAc)$_2$ precursors). All hybrid materials contain ~20 wt % CNT. The Ni/Fe precursor ratio of 5/1 was optimized for best OER activity, but similar OER catalytic activity could be obtained over the range of Ni/Fe ratio from 3/1 to 10/1.

Electrodeposition of NiFe oxides on glassy carbon: NiFe oxides were cathodically electrodeposited onto glassy carbon under the current density (1 mA/cm$^2$) for 600 second in 0.09 M Ni(NO$_3$)$_2$ and 0.01 M Fe(NO$_3$)$_3$ solution (pH=2) (the amount of catalyst is estimated to be ~2.4 mg/cm$^2$)[16].

Sample Preparation and Materials Characterizations: Scanning electron microscopy (SEM) samples were prepared by drop-drying the aqueous suspensions onto the silicon substrate, and SEM analysis was measured by an FEI XL30 Sirion scanning electron microscope. Transmission electron microscopy (TEM) samples were prepared onto copper grids by drop-drying their ethanol suspensions, and TEM characterization was carried out on an FEI Tecnai G2 F20 transmission electron microscope. XANES measurements were carried out at the SGM beamline of the Canadian Light Source. Powder samples were held onto the indium foil. XANES were measured in the surface sensitive total electron yield (TEY). Data were first normalized to the incident photon flux I0 measured with a refreshed gold mesh prior to sample measurement and then further normalized to the edge jump between pre-edge platform and post-edge platform.

Energy Dispersive Spectroscopy (EDS) analysis was carried out on Tecnai TEM with EDS detector under scanning unit electron microscopy (STEM) mode. X-ray photoelectron spectroscopy (XPS) samples were drop-dried onto silicon substrate and XPS measurement was performed on an SSI S-Probe XPS Spectrometer. X-ray diffraction (XRD) samples were prepared by drop-drying aqueous solution onto glass slides to form thick films, and XRD measurement was performed on a PANalytical X'Pert instrument. Atomic force microscopy (AFM) sample was prepared onto a silicon substrate by drop-drying the aqueous solution of the sample. AFM measurement was done with a Vecco IIIa Nanoscope in the tapping mode.

Sample Preparation for Electrochemical Characterizations: For measurements on RDE, 4 mg of catalyst was dispersed in 768 μl of water, 200 μl of ethanol and 32 μl of 5 wt% Nafion solution by at least 30 min sonication to form a homogeneous ink. Then 10 μl of the catalyst ink (containing 40 μg of catalyst) was loaded onto a glassy carbon electrode of 5 mm in diameter (loading 0.20 mg/cm$^2$). For the preparation of the carbon fiber paper electrode (from Fuel Cell Store), the hybrid catalyst was dispersed in ethanol to achieve a concentration of 1 mg/ml with 4 wt% PTFE (from its 60 wt% water suspension, Aldrich). After sonication for 30 minutes, 250 μl of the catalyst ink was drop-dried onto a 1 cm×1 cm carbon fiber paper (loading 0.25 mg/cm$^2$).

Electrochemical Characterizations: Electrochemical studies were carried out in a standard three electrode system controlled by a CHI 760D electrochemistry workstation. Catalyst powders cast on the rotating disk electrode (RDE) or carbon fiber paper was used as the working electrode, graphite rod the counter electrode and saturated calomel electrode as the reference electrode. The reference was calibrated against and converted to reversible hydrogen electrode (RHE). Linear sweep voltammetry was carried out at 5 mV/s for the polarization curves and 0.1 mV/s for Tafel plots. The hybrid catalyst was cycled ~50 times by cyclic voltammetry (CV) until a stable CV curve was developed before measuring polarization curves of NiFe-LDH/CNT. All polarization curves were corrected with 95% iR-compensation. The impedance (R) was very consistent at multiple potential points (covering both non-OER condition and OER-condition). Chronopotentiometry was carried out under a constant current density of 2.5 mA/cm$^2$ for RDE and 5 mA/cm$^2$ for CFP. Experiments involving RDE were conducted with the working electrode continuously rotating at 1600 rpm to get rid of the oxygen bubbles.

Turnover frequency (TOF) calculation of the catalysts: The TOF value is calculated from the equation[20, 37]:

$$TOF = \frac{J \times A}{4 \times F \times m}$$

$J$ is the current density at overpotential of 0.3 V in A/cm$^2$. $A$ is the area of the carbon fiber paper electrode. $F$ is the faraday constant (a value of 96485 $C/mol$). $m$ is the number of moles of the active materials that are deposited onto the carbon fiber paper.

Gas Chromatography measurement: OER catalyses were performed in a gas-tight two-compartment electrochemical cell with 1 M KOH electrolyte and Ag/AgCl reference electrode. The working electrodes were prepared by drop-drying 1 mg NiFe-LDH/CNT hybrid into 1 cm$^2$ Ni foam and 1 mg IrO$_2$ powder (from Premetek Co.) onto 1 cm$^2$ Ti plate. The wires were stamped with Ni foam and glued onto Ti plate by epoxy to ensure good connection. Chronopotientiometry was applied with different current density to maintain constant oxygen generation. Meanwhile, N$_2$ was constantly purged into the cathodic compartment with a flow rate of 5 cm$^3$/min and the compartment was connected to the gas-sampling loop of a gas chromatograph (SRI 8610C). A thermal conductivity detector (TCD) was used to detect and quantify the oxygen generated.

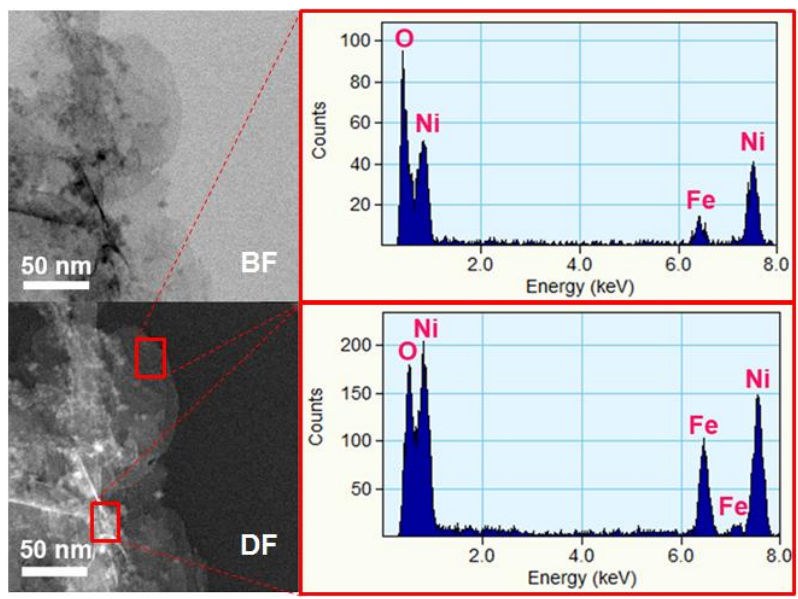

**Figure S1.** Bright field (BF) and dark field (DF) scanning transmission electron microscopy (STEM) images of NiFe-LDH/CNT hybrid. Areas in the DF images were analyzed by selected area energy dispersive spectrum (EDS) with a spatial resolution of ~2 nm. The Ni/Fe atomic ratio in the NiFe-LDH nanoplate region was estimated to be 5/1 (top right).

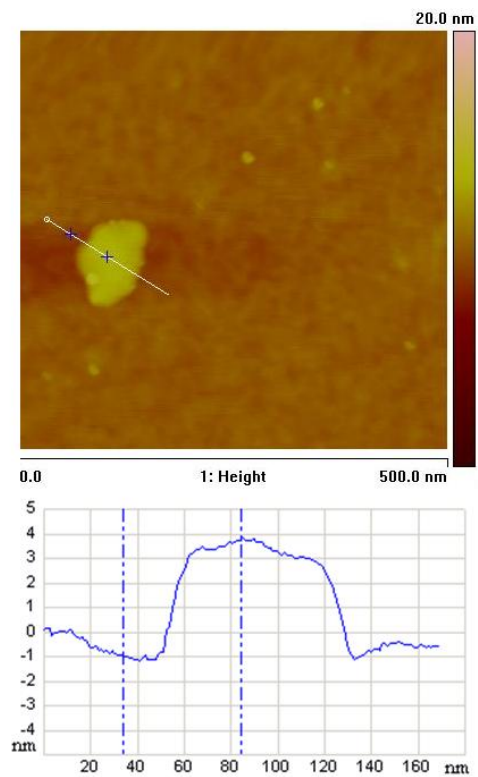

**Figure S2.** Atomic force microscopy (AFM) image and height profile of a single NiFe-LDH nanoplate. The nanoplate is ~60 nm in width and ~4 nm in thickness.

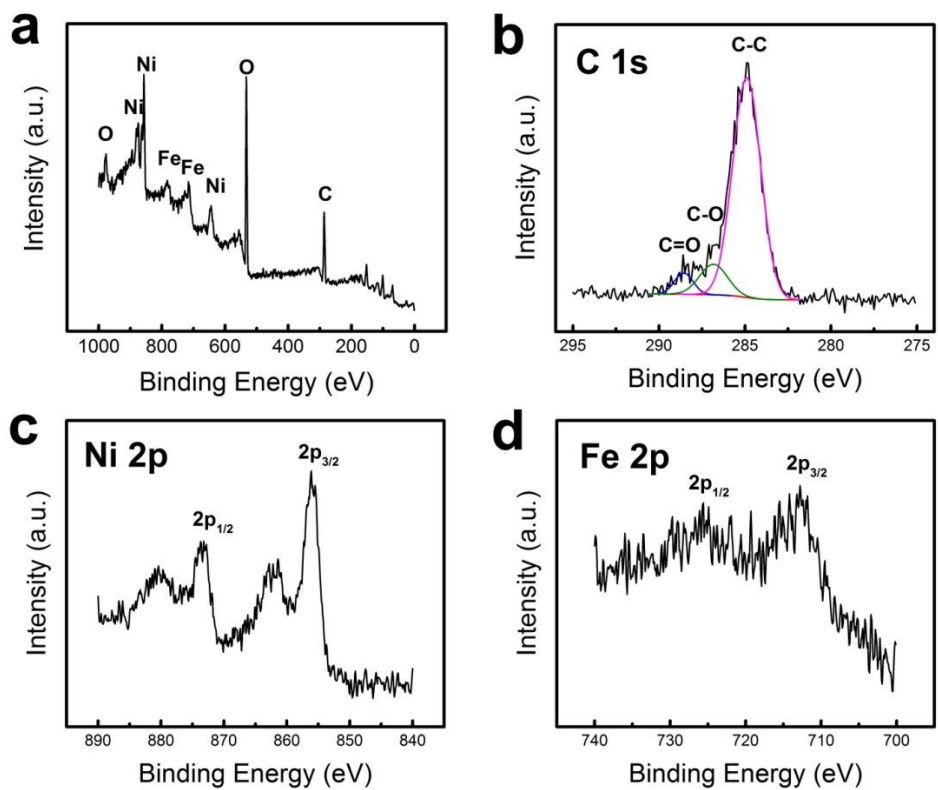

**Figure S3.** (a) XPS survey spectrum, (b) high-resolution C 1s (c) Ni 2p and (d) Fe 2p spectra of NiFe-LDH/CNT hybrid material.

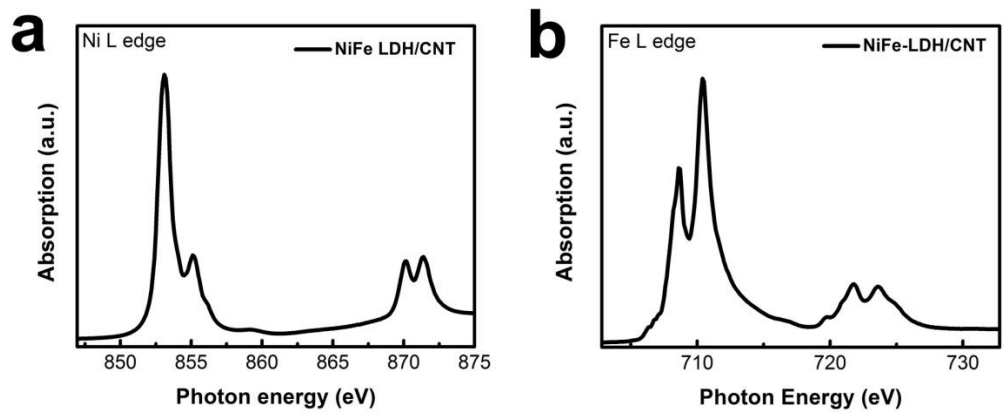

**Figure S4.** (a) Ni L edge XANES and (b) Fe L edge XANES of the NiFe-LDH/CNT hybrid.

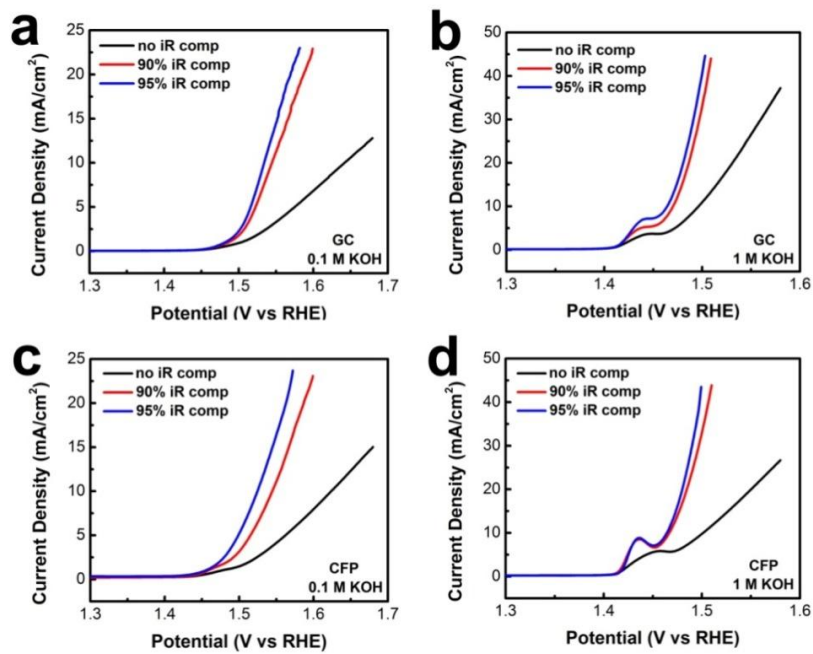

**Figure S5**. Polarization curves of NiFe-LDH/CNT under different iR compensation levels on GC in (a) 0.1 M KOH (~50 ohms) (b) 1 M KOH (~6 ohms) and on CFP in (c) 0.1 M KOH (~ 10 ohms) and (d) 1 M KOH (~ 3 ohms). 95% iR compensation is an optimal compensation level to achieve best curve shape as well as avoid curve distortion by over-compensation. The peak shift in Figure S5d caused by iR compensation is mainly due to higher absolute value of Ni redox current.

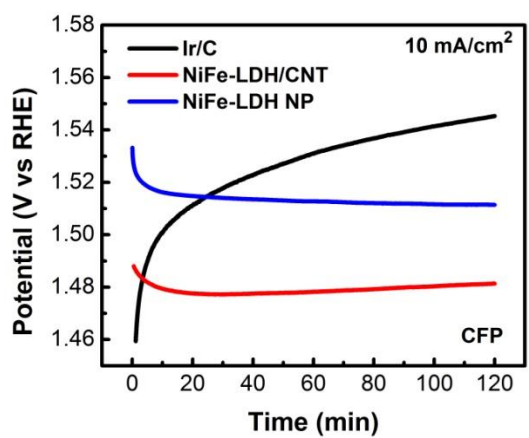

**Figure S6.** Chronopotentiometry curves of Ir/C (black), NiFe-LDH/CNT (red) and NiFe-LDH nanoplates (blue) under maximal loading in 1 M KOH at a current density of 10 mA/cm$^2$. Maximal loading of Ir/C

on CFP showed better activity than NiFe-LDH/CNT within ~3 min, but it quickly decayed and underperformed NiFe-LDH/CNT by ~60 mV after 2 hour OER catalysis. The NiFe-LDH nanoplate alone showed decent OER catalytic activity (~40 mV worse than NiFe-LDH/CNT hybrid material) and good stability (no decay over 2 hours).

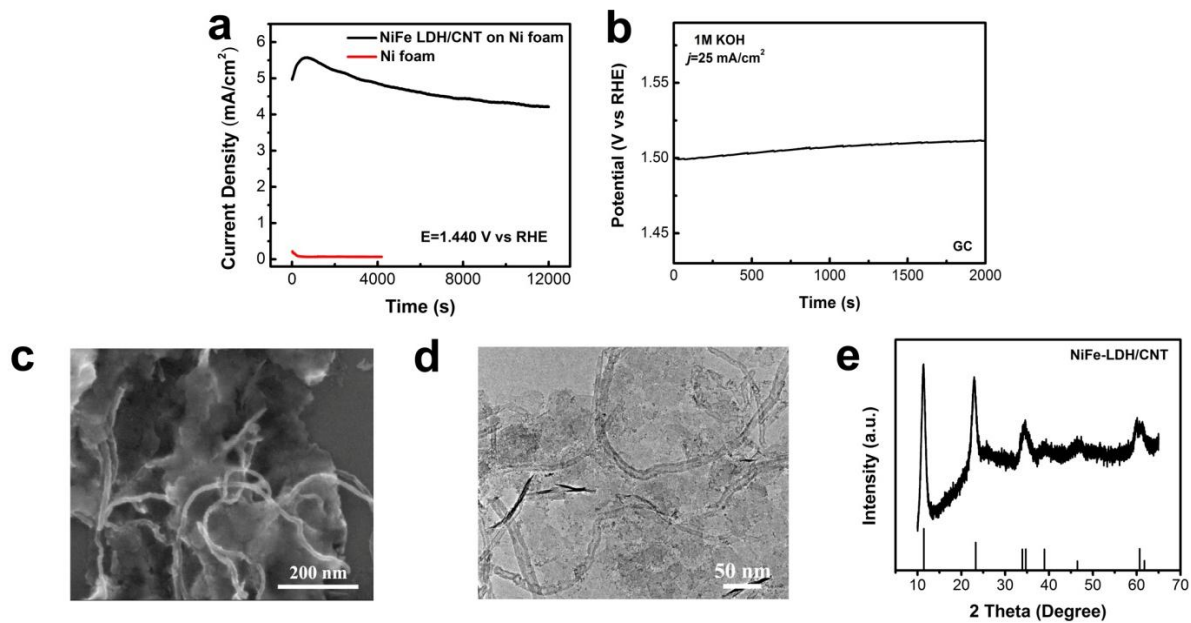

**Figure S7.** (a) Chronoamperometry curves of 3 mg of NiFe-LDH/CNT loaded into 1cm$^2$ of Ni foam (black) vs. blank Ni foam control (red) in 1 M KOH at E = 1.440V vs. RHE. Total charge passed was 56.62 Coulombs, about 3 times more than needed to full oxidize CNTs present in the hybrid. (b) Chronopotentiometry curve of NiFe-LDH/CNT on a GC electrode in 1 M KOH at the current density of 25 mA/cm$^2$ (with a loading of 0.2 mg/cm$^2$). Total charge passed was 10 Coulombs, about 20 times more than needed to fully oxidize CNTs present in the hybrid (c) SEM image and (d) TEM image of the catalyst after long time OER measurement as shown in (a). (e) XRD pattern of NiFe-LDH/CNT catalyst after anodization at 1.48 V vs RHE for 3 hours. No obvious morphological or structural transformation could be observed during OER catalysis.

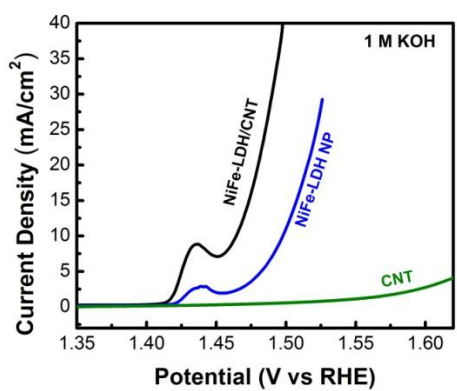

**Figure S8.** Polarization curves of pure CNT, NiFe-LDH plates alone and NiFe-LDH/CNT hybrid catalyst loaded on CFP (with a loading of 0.25 mg/cm$^2$) in 1 M KOH. Negligible CNT oxidation current could be observed OER catalysis potentials of NiFe-LDH/CNT and NiFe-LDH nanoplate.

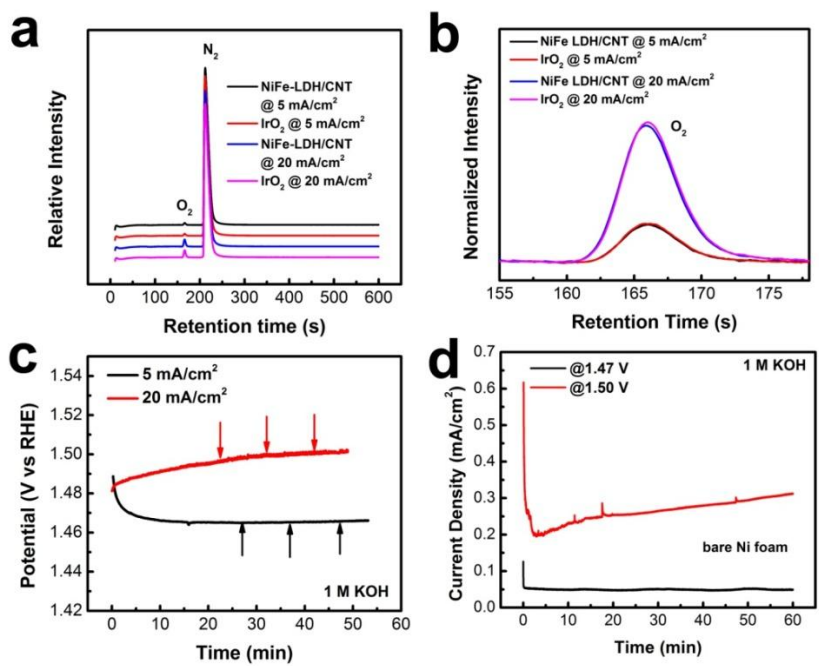

**Figure S9.** (a) Gas chromatography curves of the gaseous products from the OER catalyzed by NiFe-LDH/CNT loaded into Ni foam and commercial IrO$_2$ catalyst (IrO$_2$ powder from Premetek Co.) loaded onto Ti plate at the current density of 5 mA/cm$^2$ and 20 mA/cm$^2$. N$_2$ and O$_2$ peak position were assigned by standard gas reference. (b) Zoomed-in gas chromatography curves in oxygen peak region. The O$_2$ peaks were normalized by the relative intensity of N$_2$ peaks. The relative Faradaic efficiency of NiFe-

LDH/CNT to IrO$_2$ under 5 mA/cm$^2$ and 20 mA/cm$^2$ was calculated to be ~99.3 % and ~99.2 % respectively. Note that even though IrO$_2$ without C additives was relatively low in activity, it is free of possible carbon oxidation during OER and is a well-known standard catalyst with nearly 100% Faradaic efficiency[39], useful for calibration purposes. (c) Chronopotentiometry curves of NiFe-LDH/CNT loaded on Ni foam at a loading of 1 mg/cm$^2$ during gas chromatography measurement. The arrows point out the time-line of gas sample injection. (d) Chronoamperometry curves of bare Ni foam in 1 M KOH at the potential of 1.47 V (black) and 1.50 V (red) vs RHE. Bare Ni foam can only deliver ~0.05 mA/cm$^2$ and ~0.25 mA/cm$^2$ at the potential of which NiFe-LDH/CNT can deliver OER current densities of 5 mA/cm$^2$ and 20 mA/cm$^2$ respectively.

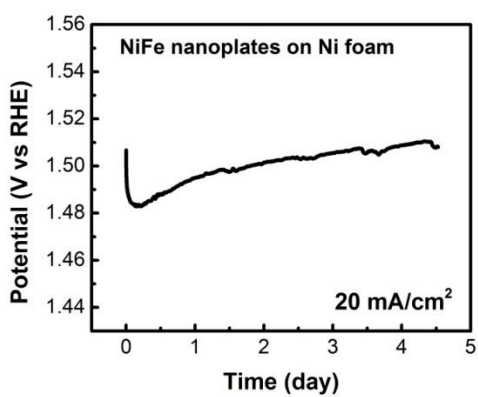

**Figure S10.** Chronopotentiometry curves of free NiFe LDH nanoplates (loading of ~5 mg/cm$^2$) in 1 M KOH loaded into Ni foam at a current density of 20 mA/cm$^2$ to evolve oxygen over > 4 days.

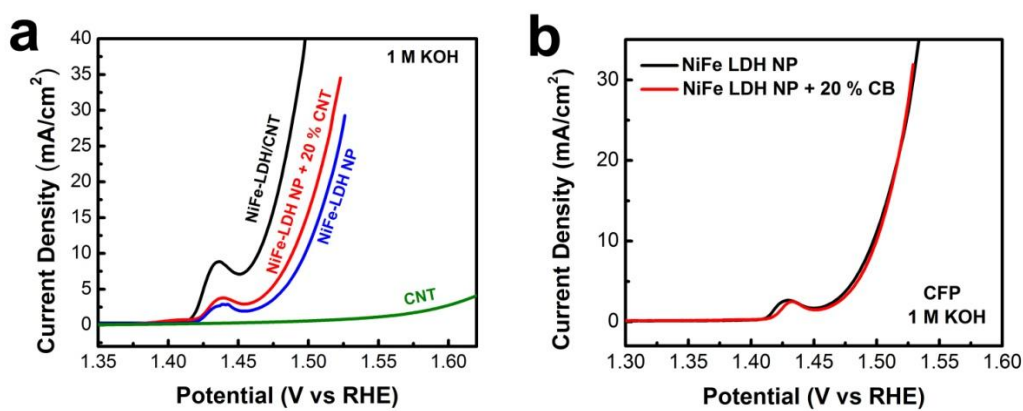

**Figure S11.** a) Polarization curves of NiFe-LDH/CNT (black), NiFe-LDH/CNT mixed with 20% CNT (red), NiFe-LDH nanoplate (blue) and CNT alone (green) on CFP (with a loading of 0.25 mg/cm$^2$) in 1 M KOH. b) Polarization curves of NiFe-LDH/CNT (black) and NiFe-LDH/CNT mixed with 20% carbon black (CB) (red) on CFP (with a loading of 0.25 mg/cm$^2$) in 1 M KOH

**Table S1.** OER activities of some benchmark catalysts in alkaline solution

| Material | Electrolyte | Current Density | Over-Potential | Reference |
|---|---|---|---|---|
| NiFe-LDH/CNT | 0.1 M KOH | 10 A/g | 1.511V | This work* |
| | | 10 mA/cm$^2$ | 1.538V | This work |
| | 1 M KOH | 10 A/g | 1.458V | This work |
| | | 10 mA/cm$^2$ | 1.477V | This work |
| RuO$_2$ | 0.1 M KOH | 10 A/g | 1.528V | Ref. 8 |
| IrO$_2$ | 0.1 M KOH | 10 A/g | 1.518V | Ref. 8 |
| Ba$_{0.5}$Sr$_{0.5}$Co$_{0.8}$Fe$_{0.2}$O$_3$ | 0.1 M KOH | 10 A/g | 1.53V | Ref. 15 |
| Mn$_3$O$_4$/CoSe$_2$ | 0.1 M KOH | 10 mA/cm$^2$ | 1.68V | Ref. 38 |
| Co$_3$O$_4$/N-rmGO | 1 M KOH | 10 mA/cm$^2$ | 1.54V | Ref. 11 |
| core-ring NiCo$_2$O$_4$ | 1 M KOH | 100 mA/cm$^2$ | 1.545V | Ref. 13 |
| NiFe(OH)$_2$ | 1 M NaOH | 500 mA/cm$^2$ | 1.495V (353K) | Ref. 19 |
| NiFe oxide | 1 M NaOH | 0.5 mA/cm$^2$ | 1.51V | Ref. 23 |

*Determined by Tafel plot and polarization curve

## Additional References